\documentclass[useAMS,usenatbib]{mn2e}
\usepackage{graphicx,mathrsfs}

\makeatletter

\newcommand{\Rmnum}[1]{\expandafter\@slowromancap\romannumeral #1@}
\makeatother

\title[Photo-$z$ Performance for Precision Cosmology \Rmnum{2}]{\begin{center}
Photo-$z$ Performance for Precision Cosmology \Rmnum{2} : Empirical Verification\thanks{Based on observations undertaken at the European Southern Observatory (ESO) Very Large Telescope (VLT) under Large Program 175.A-0839. Also based on data collected at Subaru Telescope, which is operated by the National Astronomical Observatory of Japan.}\\
\end{center}}

\author[R.~Bordoloi et al.]
{\parbox{\textwidth}{R.~Bordoloi,$^{1}$\thanks{E-mail: \texttt{rongmonb@phys.ethz.ch}}
S. J. ~Lilly,$^{1}$
A.~Amara,$^{1}$ 
P. A.~Oesch,$^{1,2}$ 
S.~Bardelli,$^{3}$
E.~Zucca,$^{3}$
D.~Vergani,$^{4,3}$
T.~Nagao,$^{5,7}$
T.~Murayama,$^{6}$
Y.~Shioya$^{5}$ \&
 Y.~Taniguchi$^{5}$}\vspace{0.4cm}\\
\parbox{\textwidth}{$^{1}$Institute for Astronomy, ETH Z\"{u}rich, Wolfgang-Pauli-Strasse 27, CH-8093, Z\"{u}rich\\
$^{2}$UCO/Lick Observatory, Department of Astronomy and Astrophysics, University of California, Santa Cruz, CA 95064\\
$^{3}$INAF Osservatorio Astronomico di Bologna, Bologna, Italy\\
$^{4}$Istituto Nazionale di Astrofisica - Istituto di Astrosica, Spaziale e Fisica Cosmica Bologna, via P. Gobetti 101, I-40129 Bologna, Italy\\
$^{5}$Research Center for Space and Cosmic Evolution,Ehime University,Bunkyo-cho, Matsuyama 790-8577, Japan\\
$^{6}$Astronomical Institute, Graduate School of Science, Tohoku University, Aramaki, Aoba, Sendai 980-8578, Japan\\
$^{7}$The Hakubi Project, Kyoto University, Yoshida-Ushinomiya-cho, Sakyo-ku, Kyoto 606-8302, Japan\\}}

\begin{document}

\date{Accepted ---- . Received ----; in original form ----}

\pagerange{\pageref{firstpage}--\pageref{lastpage}} \pubyear{----}

\maketitle

\label{firstpage}

\begin{abstract}
The success of future large scale weak lensing surveys will critically depend on the accurate estimation of photometric redshifts of very large samples of galaxies. This in turn depends on both the quality of the photometric data and the photo-$z$ estimators.  In a previous study, \citep{Bordoloi09} we focussed primarily on the impact of photometric quality on photo-$z$ estimates and on the development of novel techniques to construct the $N(z)$ of tomographic bins at the high level of precision required for precision cosmology, as well as the correction of issues such as imprecise corrections for Galactic reddening.  We used the same set of templates to generate the simulated photometry as were  then used in the photo-$z$ code, thereby removing any effects of ``template error''.  In this work we now include the effects of ``template error'' by generating simulated photometric data set from actual COSMOS photometry. We use the trick of simulating redder photometry of galaxies at higher redshifts by using a bluer set of passbands on low $z$ galaxies with known redshifts.   We find that ``template error'' is a rather small factor in photo-$z$ performance, at the photometric precision and filter complement expected for all-sky surveys.  With only a small sub-set of training galaxies with spectroscopic redshifts, it is in principle possible to construct tomographic redshift bins whose mean redshift is known, from photo-$z$ alone, to the required accuracy of $|\Delta_{< z >}| \leq 0.002(1+z)$.
\end{abstract}

\begin{keywords}
galaxies: distances and redshifts- cosmology: observations- methods: Statistical
\end{keywords}

\section{Introduction}

In the coming years, the next generation of all sky surveys plan to map the whole extragalactic sky in visible \& near infra-red photometric bands. In order to achieve the performance required for precision cosmology, such surveys will have a depth of $ I_{AB} \leq 24.5$, and contain multi-band photometric measurements of the order of  $\sim 10^{9}$ galaxies over a $ 2 \upi$ sr survey area. The use of these surveys for precision cosmology and for many other types of science will require the use of photometrically-estimated redshifts (hereafter photo-$z$).  The use of photo-$z$ for precision cosmology, e.g. in weak-lensing tomography \citep{Hu_Tomography}, will require an impressive performance from the photo-$z$, not only in statistical accuracy per object, but especially in the control of systematic biases in the mean redshifts of sets of galaxies used for cosmological analysis.  It is easy to derive that a precision of 1\% in the dark energy equation of state parameter ($w$), requires knowledge of the redshifts of the probes with a systematic accuracy of $0.002(1+z)$.

In a previous paper \citep{Bordoloi09} (hereafter BLA-10), we simulated the photometric redshift performance for a number of different photometric depths.  These were defined as a combination of a Euclid-like infrared survey \citep{EIC2010}, complemented with additional ground-based photometry in the visible waveband.  Artificial photometric catalogues were generated from the mock COSMOS light-cone catalogues of \cite{kitzbichler&White2007} and photo-$z$ of galaxies were produced using a template fitting code ZEBRA \citep{Feldmann2006}.  We deliberately used exactly the same set of  templates to construct the simulated photometry as were then used in ZEBRA to estimate the photo-$z$.  This choice eliminated the effects of ``template-error'' on the photo-$z$, by construction, allowing us to focus on other issues.  We argued in BLA-10 that the exceptional performance of photo-$z$ in COSMOS (see \citealt{2009ApJ...690.1236I}), where deep photometry is available in a large number of filters, suggested that template error was unlikely to be a dominant factor with the poorer and more limited photometric data that can realistically be anticipated over the whole sky.  However, that analysis nevertheless represents an idealized situation, and a certain ``circularity'' of the argument.  In this paper we aim to rectify this shortcoming by basing the simulated photometry on actual photometric data.

Clearly, the ideal approach would be to take already available photometry that was comparable to that which will be available for the all-sky surveys, estimate the photo-$z$, and compare the performance with a complete set of reliable spectroscopically-determined redshifts.  However, at present, there is no photometric survey that gives near-infrared photometry at the depth envisaged for e.g. Euclid ($Y,J,H = 24$ at 5$  \sigma$ \citep{EIC2010}), nor is there a sufficiently complete spectroscopic sample that extends down to $I_{AB} \leq 24.5$, where the redshift distribution extends well into the ``redshift desert'' at $z > 1.2$.  

We can however perform the following trick. To first order, the photo-$z$ performance that we are interested in, i.e. on galaxies at some redshift using some given set of filters at some photometric depth, can be simulated using photometry on a set of brighter galaxies at slightly lower redshifts using a bluer set of filters. These photo-$z$ can also then be tested on a brighter spectroscopic sample at more benign (lower) redshifts.  We argue below that degraded $uBVrizJ$ photometric data from COSMOS, plus high quality redshifts from the zCOSMOS-bright redshift survey \citep{Lilly2007,Lilly2009_article} at $V \leq 22.5$ and $0.2 < z < 1.0$ are well-matched (including in intrinsic luminosity) to a survey at $0.7 < z < 1.8$ with $I_{AB} \leq 24.5$, i.e. with a change in $(1+z)$ of a factor $\sim 1.4$.  Of course, the process is not perfect: it might be that templates (even at similar luminosities) become pathological between $0.2 < z < 1.0$ and $0.7 < z < 1.8$ but there is in our view not much evidence to support this.  Perhaps more seriously, the number density of galaxies will be artificially low, by a factor of about four, so the problem of overlapping objects that we discussed in BLA-10 will not be fully accounted for.

The paper is organized as follows: in Section 2 we describe the photometric and spectroscopic catalogues that are used and the construction of the simulated photometric catalogue. Then we proceed to describe in Section 3 how the photo-$z$ are estimated and finally in Section 4 how they perform.  Much of the performance analysis follow closely the approach of BLA-10 to which the reader is referred for more details.

\begin{figure}
\includegraphics[angle=0,width=8cm,height=7cm]{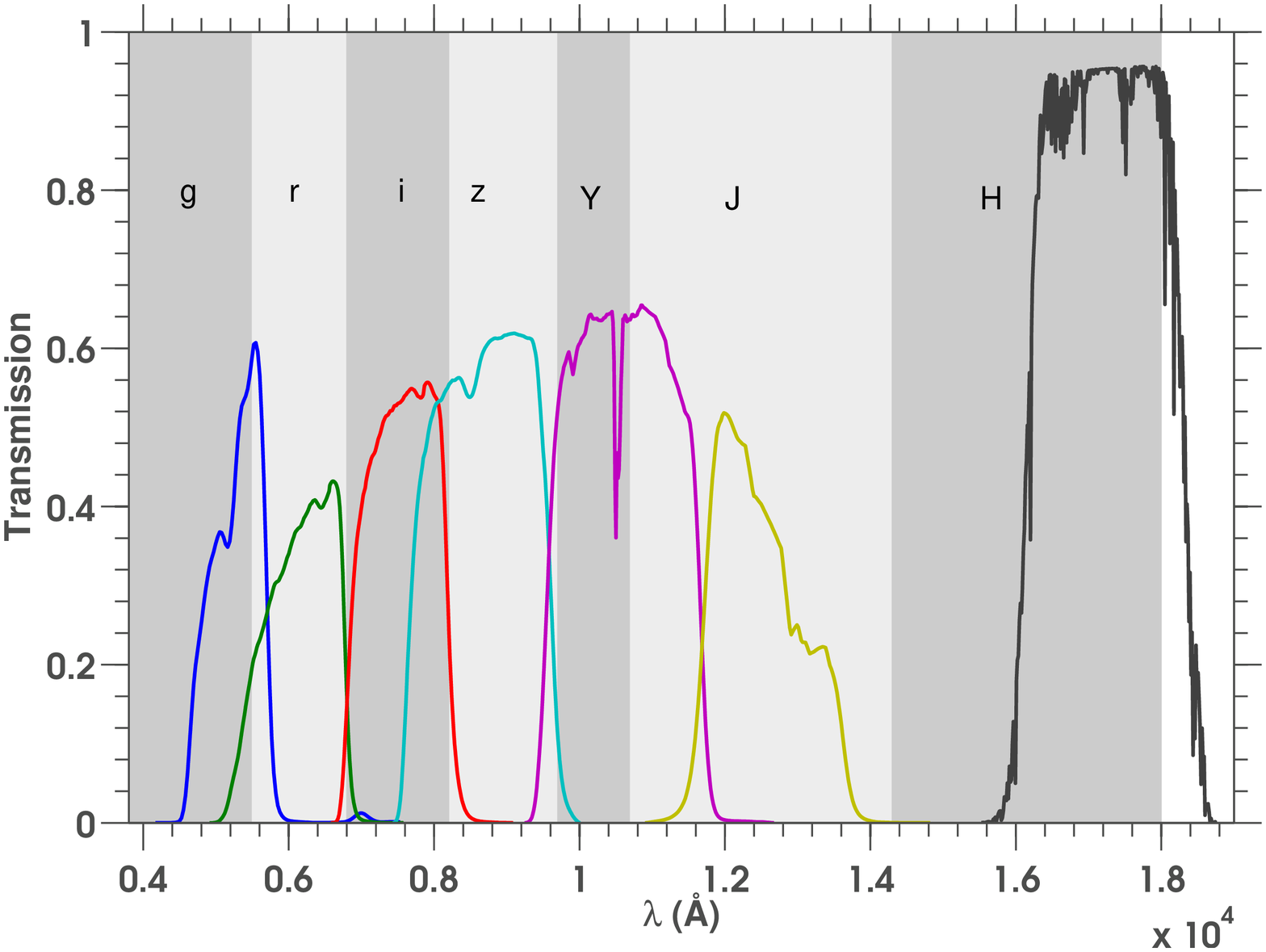}

\caption{Actual COSMOS filter response curves, after transformation by a factor of 1.38 in wavelength (as described in Section 2.3), compared with the nominal top-hat filters used in BLA-10 to simulate an all-sky survey combining ground- and space-based data.  There is a reasonable match in spacing and width.  Since the current photo-$z$ analysis used these transformed filter response curves, the differences should not be important.}
\label{Transformed_filters}
\end{figure}

\section{Creating the mock catalogue}

Our goal is to simulate a realistic photometric survey consisting of visible photometry of the depth and filter coverage of planned surveys.  In BLA-10, we defined three nominal surveys that were defined by increasing photometric depth in the visible, combined with NIR photometry of the depth expected from the proposed Euclid survey. The second of these, Survey B, which we argued had just sufficient performance, was intended to simulate the depth of the Pan-STARRS-2\footnote{http://pan-starrs.ifa.hawaii.edu} or DES\footnote{http://www.darkenergysurvey.org} surveys.  For precise comparison with the results of BLA-10, we use this same intermediate depth Survey B for the present analysis.

\subsection{Input photometric data}

We use the COSMOS photometric catalogue which contains fluxes measured in 30 bands from Subaru, HST, CHFT, UKIRT, Spitzer and GALEX telescopes \citep{2009ApJ...690.1236I}. We refer the reader to \cite{cosmos_photometry_catalogue} for a complete description of observations, data reductions and the photometry catalogue. For our present purpose we use only the $uBVrizJ$ broad-band filter data from this catalogue as described in detail in section 2.3 .

\subsection{Input spectroscopic data}

We use the spectroscopic data from the zCOSMOS survey \citep{Lilly2007}. We use the so-called ``20k-bright'' sample of 20,000 galaxies selected at $I_{AB} \leq 22.5$ (Lilly et al., in preparation). These galaxies were observed with the VIMOS/VLT spectrograph covering a wavelength range of $5500 \AA \leq \lambda \leq 9000 \AA$ at a resolution of 600. The average accuracy of individual redshifts is 110 km $\rm{s^{-1}}$, irrespective of redshift.  

Unfortunately, the spectroscopic sample is not 100\% complete. As described in \citealt{Lilly2009_article}, each redshift in zCOSMOS is assigned to a confidence class that is based on both a subjective assessment of the reliability of the spectroscopic measurement and whether there is a broad consistency between the spectroscopic and photometric redshifts, i.e. whether $|\Delta z| < 0.08(1+z)$.  The photo-$z$ come from from the full COSMOS photometric data set, as described in \citealt{2009ApJ...690.1236I}.  The actual reliability of the redshifts across the confidence class scheme is estimated using the repeatability of the redshifts from duplicated spectroscopic observations. In the current work we only make use of the redshift measurements having confidence classes ( 4.x, 3.x, 9.5 + 9.4 + 9.3 + 2.5 + 2.4 + 1.5) which are demonstrated from the repeat observations to be 99 \% reliable \citep{Lilly2009_article}.  The decimal places 
in the confidence class say whether there is broad agreement between the spectroscopic and photometric redshifts, e.g. a decimal place of .5 signifies agreement between the spectroscopic and photometric redshift estimates for that object, otherwise they do not agree. The confidence classes marked 4.x and 3.x are objects with very secure spectroscopic redshifts irrespective of their photometric redshifts.

As detailed below, we restrict the redshift range of the sample to $0.2 < z < 1.0$ and select galaxies to have $V \leq 22.5$.   We exclude stars and broad line AGN from their spectroscopic flags (as they are point sources, they can be photometrically identified and will not be used in a weak lensing analysis). Reliable redshifts (as defined above) are available for 83.5$\%$ of the remaining galaxies. Of these, 80\% have very secure  redshifts (99.5\% reliable) based on the spectra alone. The remainder have less secure spectroscopic redshifts that are trusted only because they are consistent with the photo-$z$.  Clearly, any of these galaxies which have a pathological spectral energy distribution yielding a wrong photo-$z$ will have been excluded from our sample.  However, any such pathological objects for which we secured a reliable redshift from the spectrum alone would be included and would be a ``catastrophic failure'' in the photo-$z$.  It should be noted that the most secure spectro-$z$ have a catastrophic failure rate in photo-$z$ of about 1\%.  

It should be recognized that this procedure, which we cannot avoid, means that the current analysis may underestimate the number of ``catastrophic failures'' in the photo-$z$, and thus the photo-$z$ quality, in the overall sample if the incidence of failure, or the photo-$z$ error, are in some way correlated with the difficulty of measuring a purely spectroscopic redshift \citep{Hildebrandt2008}.  As discussed in \cite{Lilly2007} there is in fact a very good correspondence between the spectroscopic repeatability of the redshifts and the fraction of objects with consistent photo-$z$, indicating that the vast majority of the objects with less secure spectroscopic redshifts that are inconsistent with the photo-$z$, are in fact wrong spectroscopic identifications.   If the fraction of real ``catastrophic failures'' in the photo-$z$ is not correlated with the success in measuring a spectroscopic redshift, then the number of such failures would be under-estimated in the current sample by 20\%, i.e. of order 0.2\% in the overall sample, which we believe is unimportant for the current purposes.

This point does however emphasize the need to have large spectroscopic samples in the future with very high completeness in securing highly reliable redshifts if photo-$z$ schemes are to be adequately tested and characterized at the levels of precision required.  With this caveat in mind, the present purposes are nevertheless served by the available spectroscopic sample.

\begin{figure}
\includegraphics[angle=0,width=8cm,height=7cm]{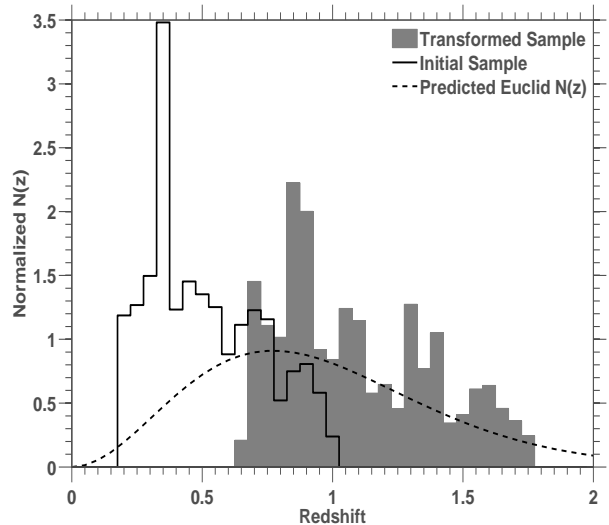}

\caption{The redshift distributions of the samples under study. The solid line gives the normalized distribution of the original set of zCOSMOS galaxies within $0.2 \leq z \leq 1.0$ and $V \leq 22.5$, as described in the text. The shaded histogram gives the normalized distribution of the same set of galaxies after transforming to higher redshift using equation 1, spanning $0.65 < z < 1.8$. For comparison, the dashed line gives the normalized predicted $N(z)$ for $I_{AB} \leq 24.5$ adopted from \citealt{Amara&Refrigier2007}}
\label{Transformed_distribution}
\end{figure}

\subsection{Generating the mock photometric catalogues}

As outlined above, we assume that bluer photometry of lower redshift galaxies (at $0.2 < z < 1.0$) can be used as a substitute for redder photometry of higher redshift galaxies.  Specifically, we take actual broad band $uBVrizJ$ photometry from COSMOS to mimic the $grizYJH$ bands for the simulated data. As shown in Figure-\ref{Transformed_filters}, this broadly corresponds to a multiplicative transformation in wavelength given by $\lambda' = \delta \lambda$, with $\delta \sim 1.38$. The match is naturally not perfect, but this should not matter since we will use the transformed COSMOS filter passbands in the photo-$z$ code and it is unlikely that the precise location of photometric bands affects the photo-$z$ performance.  In Figure-\ref{Transformed_filters} the vertical shaded regions give the wavelength ranges of the filters of Survey B, while the curves represent the broad band COSMOS filters after the above wavelength transformation. Given that the match is not perfect, the photo-$z$ performance obtained in this paper should therefore not be regarded as exact ``predictions'', but rather are intended to enable us to examine the general issue of ``template error'' in template fitting photo-$z$ codes.  

This wavelength transformation is equivalent to a redshift transformation of galaxies as follows:
\begin{equation}
\rm{z'} = \delta(1 + \rm{z})  - 1 
\label{redshift_transform}
\end{equation}

We first isolate all the galaxies in zCOSMOS-bright with secure redshifts $0.2 < z_{spec} < 1.0$ which do not lie in photometrically masked areas of \cite{2009ApJ...690.1236I} and which have complete photometric information in the selected bands.  Since essentially all $I_{AB} \leq 22.5$ galaxies have $(V-I)_{AB} > 0$ we can assume that this sample will be complete down to $V \leq 22.5$ and we retain only these galaxies.  This gives us a sample of approximately 8500 galaxies down to $V \leq 22.5$. 

To convert the available COSMOS photometry to simulate Survey B we rename the filters $uBVrizJ$ to $grizYJH$ and add 2.0 magnitudes to all the photometric measurements of each galaxy.  This creates a fainter photometric catalogue that should be complete in the ``transformed I-band'' to $I_{tr,AB} \leq 24.5 $.  The filter response curves are carried over by shifting the wavelengths by a factor 1.38.

Finally we apply the above redshift transformation to all the spectroscopic redshifts. This produces a new redshift distribution between $0.65 < z < 1.8$ that is well matched to that expected, at $z > 0.65$, for an $I_{AB} \leq 24.5$ sample in the sky.  This is shown in Figure-\ref{Transformed_distribution} where the dashed line gives the normalized predicted redshift distribution for an $I \leq 24.5$ sample, with a median redshift $ \rm{z_{median} = 0.9}$, adopted from \citealt{Amara&Refrigier2007}, compared with the redshift distribution of the original $V \leq 22.5$ zCOSMOS sample and the ``transformed'' $I_{AB} \leq 24.5$ sample. The redshift range at $z > 0.8$ is where the performance of photo-$z$ generally is less proven.  Furthermore, it should be noted that the increased distance modulus (2.2 between $z = 0.2$ and $z = 0.7$ and 1.6 between $z = 1$ and $z = 1.8$, allowing for a band-width stretching term) is well-matched to the 2.0 magnitude increase in depth between $V = 22.5$ and $I_{AB} = 24.5$.  This means that the transformed simulated galaxies are well-matched in absolute magnitude to the lower redshift input sample.

The catalogue thus created therefore contains actual photometric data from the COSMOS survey. As a result, It effectively eliminates the circularity problem of using the same SEDs to create a mock photometric catalogue and to estimate the photo-$z$ of the galaxies. In view of the +2 mag transformation of the magnitudes, the noise in the COSMOS flux densities is artificially low. We therefore add additional Gaussian noise to the transformed flux densities in each pass-band.  We use the adopted survey parameters as given in Table 1 of BLA-10 to define the required noise on each galaxy in each band, and add random noise in quadrature to the (known) uncertainties in the transformed COSMOS photometry so as to match this required final uncertainty.

Although it eliminates the obvious circularity of using templates to construct the catalogue, this procedure is clearly not perfect.  It first assumes that any problems associated with the templates are no more severe at $1.0 < z < 1.8$ than they are at $0.2 < z < 1.0$.  As noted above, the luminosities of the galaxies are rather similar at the two redshifts since the dimming corresponds quite closely to the change in distance modulus.  Changes in metallicity and in the overall specific star-formation rates are certainly present between these redshifts \citep{Lilly1996,Elbaz2007,Daddi2007a} but are not dramatic. \cite{Mannucci2010} have shown that galaxies up to $z \sim 2.5$ follow a fundamental metallicity relation (FMR) with a residual dispersion of $\sim$ 0.05 dex in metallicity( 12 \%) and with no indication of evolution with redshift.  Other more practical shortcomings are that problems associated with over-lapping objects (see BLA-10 for a discussion) will be less severe, since the parent photometric catalogue is effectively based on brighter galaxies, with a roughly fourfold lower surface density on the sky. Furthermore, the dominant uncertainties in the photometry come from our added ``artificial'' Gaussian term.  
\begin{figure}
\includegraphics[angle=0,width=8cm,height=7cm]{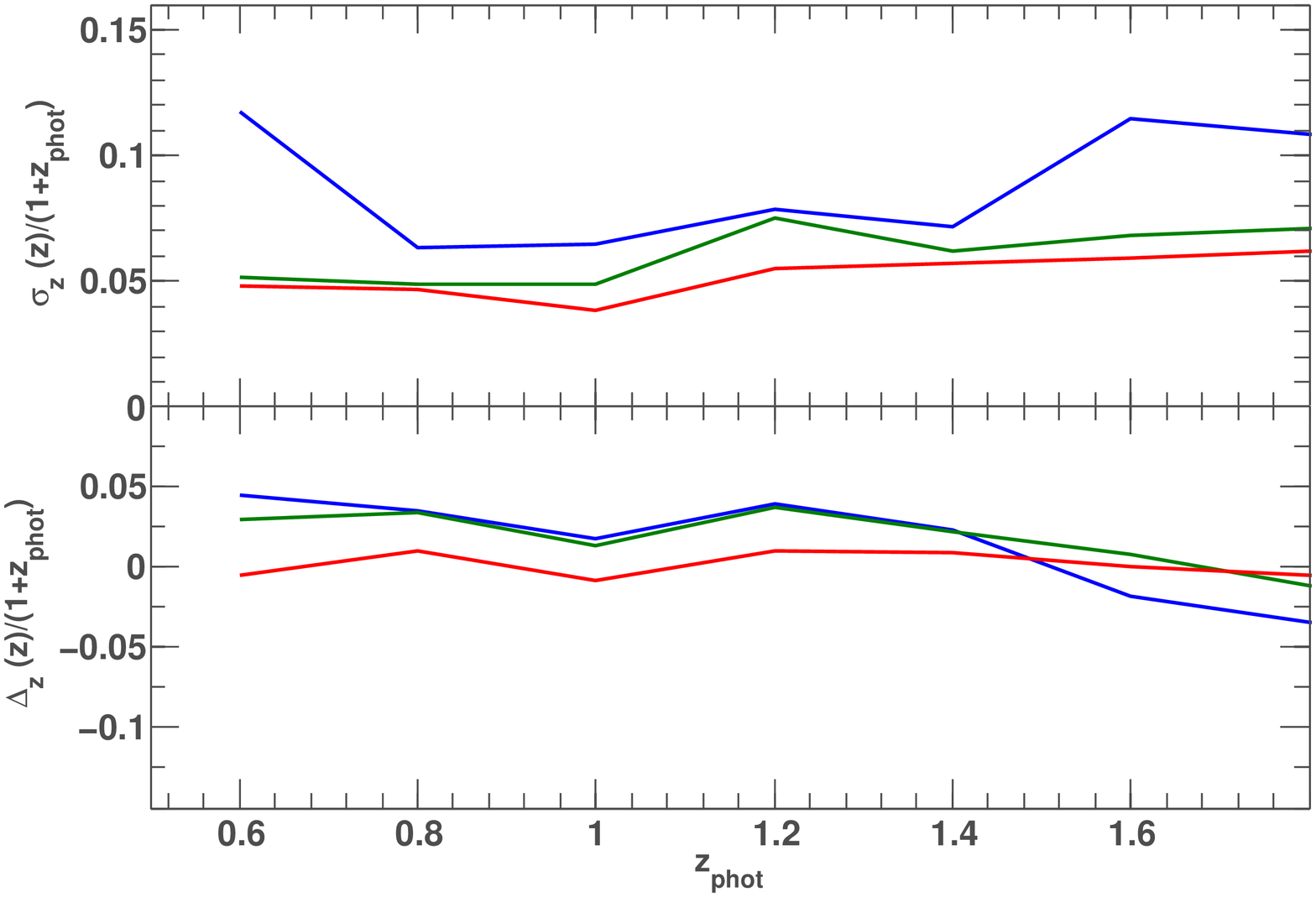}
\caption{The overall photo-$z$ performance for Survey B depth, based on a single ``maximum likelihood'' photo-$z$ for each galaxy. The blue line gives the initial performance without any cleaning, the green line after ``a priori'' cleaning, which eliminates almost 20$\%$ of galaxies, and the red line gives the performance after cleaning and after generating the probability distribution functions from the likelihoods as in \citet{Bordoloi09}.  This Figure should be compared with Figure-2 in \citet{Bordoloi09}.}
\label{overall_performance}
\end{figure}

\begin{figure}
{\includegraphics[angle=0,width=8cm,height=7cm]{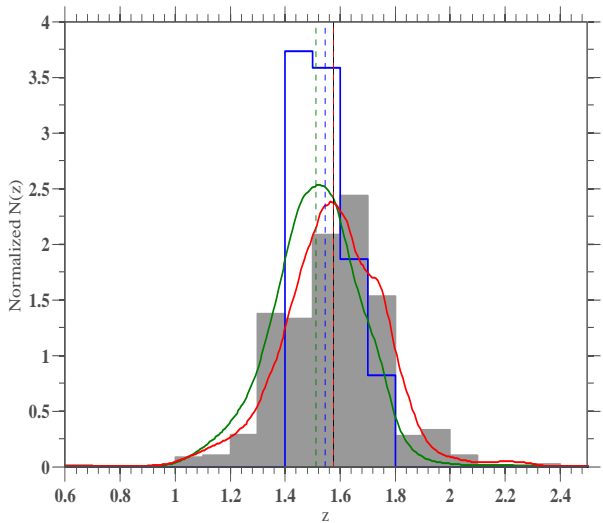}}
\caption{ Examples of reconstructed $N(z)$ for maximum likelihood redshifts (blue) after ``a priori'' cleaning of outliers, sum of the uncorrected likelihood functions (green) and sum of pdfs obtained by applying a statistical correction (red). The filled histogram represents the true $N(z)$ of the tomographic bin. The vertical lines show the mean redshift for each reconstruction. The sum of pdfs (red) represent the best reconstruction of the true $N(z)$. }
\label{tomo_bin_example}
\end{figure}

\section{Estimating the photo-z}

The estimation of photo-$z$ is done in the same way as was done in BLA-10, using the template-fitting photo-$z$ code ZEBRA \citep{Feldmann2006} to calculate photo-$z$ for the galaxies in the final simulated catalogue.  ZEBRA gives both a single best fit ``maximum likelihood'' redshift and template type, together with confidence limits based on the $\chi^{2}$ surfaces, and normalized likelihood functions $L(z)$ for the individual galaxies.   The filter response curves used in ZEBRA for this analysis are generated from the actual COSMOS filter curves, shifted in wavelength by a factor of 1.38 as described above.

We stress that the templates used in this work are purely synthetic stellar populations generated from \cite{Bruzual&Charlot2003} evolutionary models and the parameters varied to compute the whole set of templates are age, star formation history (continuous underlying SFR superimposed with single or multiple bursts), dust obscuration in the galaxy and stellar metallicity. These templates include a range of internal reddening, which ranges from $0 < A_{v} < 2$ magnitudes.  Intergalactic absorption in each template is compensated for using the Madau law \citep{Madau1995}. A final set of 10,000 templates are produced which are used to estimate photo-$z$ for the simulated survey.  The standard small adjustment of photometric zero-points was undertaken \citep{2009ApJ...690.1236I}, but we emphasize that {\it the templates were not optimized using the COSMOS photometry}.

\begin{figure*}
{\includegraphics[angle=0,width=12.5cm,height=10cm]{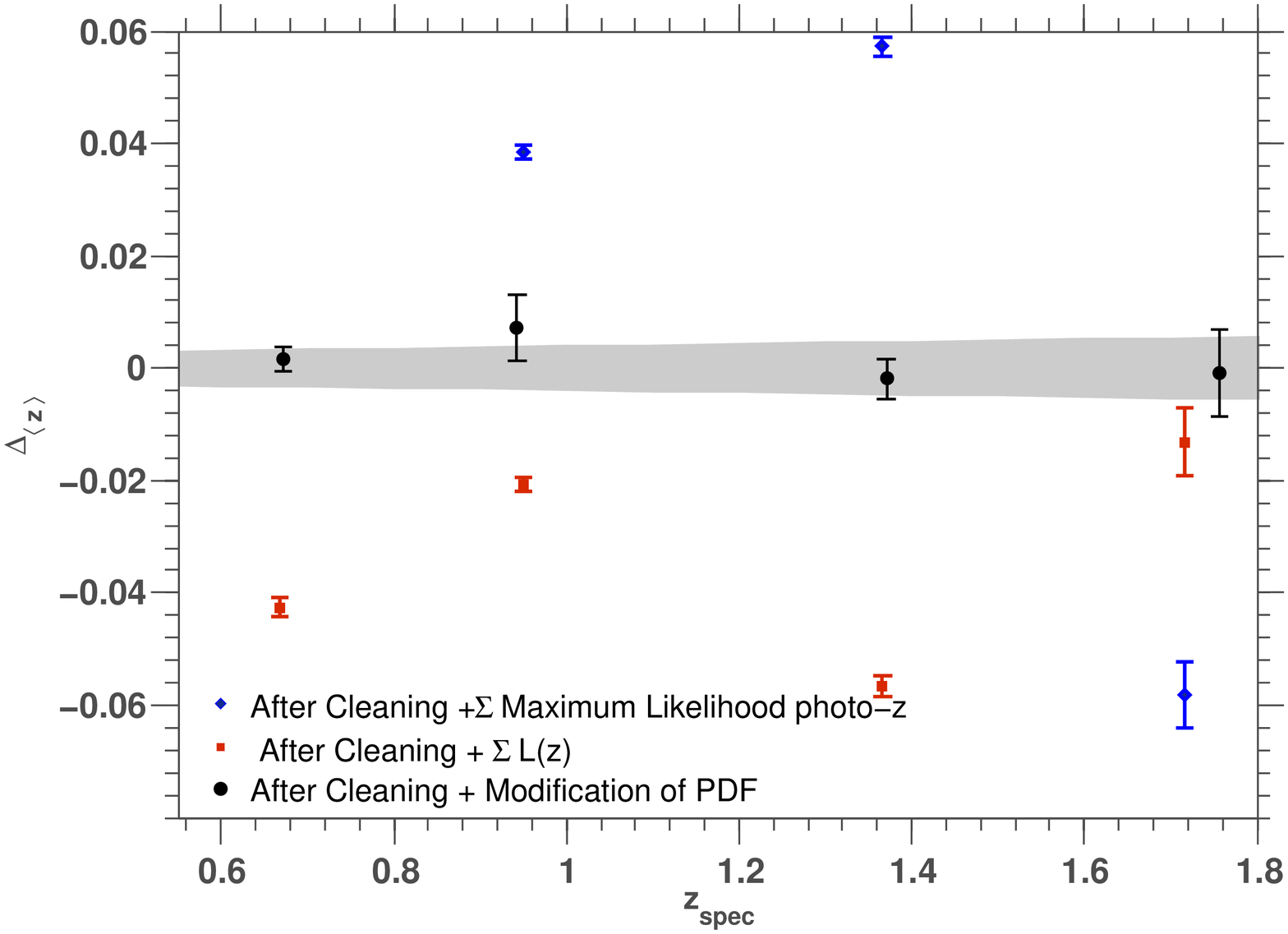}}

\caption{The error in the mean of the $N(z)$ for tomographic redshift bins for the simulated Survey B.  The blue diamond symbols are obtained using only the maximum likelihood redshifts, the open squares are obtained from the sum of the uncorrected likelihood functions, and the filled circles by summing the pdfs obtained by applying a statistical correction. All three cases are after the ``a priori'' removal of about 20\% of galaxies. The shaded region gives the nominal requirement of $|\Delta_{< z >}| \leq 0.002(1+z)$.}
\label{deltaz_final}

\end{figure*}

\section{Photo-z performance}

In this section we assess the performance of the photometric redshift estimator as applied to our more realistic simulated photometry and quantify the results in comparison with the results of BLA-10.  We follow closely the procedures of that first paper and refer the reader to it for further technical details. 

Galaxies are first binned in redshift on the basis of their observed maximum-likelihood photo-$z$.  The bias $\Delta_{z}(z)$ and dispersion $\sigma_{z}(z)$, as defined in \cite{Bordoloi09}, are computed as a function of the observed photo-$z$. As in BLA-10, the samples are cleaned by applying an ``a priori'' cleaning algorithm that is based solely on the shape of the output $L(z)$ function. The redshift probability distribution function (pdf) of each galaxy, $\mathscr{L}(z)$, is constructed from its individual likelihood function $L(z)$ using the procedure described by BLA-10. This is based on the principle that the real redshifts (for those galaxies with known redshifts) should, within a large enough sample, be uniformly distributed throughout their respective pdfs.  Some set of galaxies with known (spectroscopic) redshifts is therefore used to construct a mapping between L(z) and $\mathscr{L}(z)$ which is assumed to be the same, in cumulative probability space, for all galaxies. The reader is referred to BLA-10 for details of this algorithm, and the mapping of   $L(z)$ to $\mathscr{L}(z)$ is given by the equation 12 and shown in Figure 3 of BLA-10. In the current simulation, this transformation is constructed using a random subset of $\sim$10$\%$ of the total sample, (i.e. of the order of $10^{3}$ galaxies), the remaining 90\% then serving as a test. 

In weak lensing tomography using photo-$z$, there are two principal requirements.  First, the photo-$z$ of individual galaxies must be known to a precision of about $\sigma_z < 0.05(1+z)$ in order that tomographic bins can be defined without significant overlap, thereby reducing the effects of ``intrinsic alignments''.  Second, and more demanding, the $N(z)$ of each of these bins must be known accurately for the correct quantitative interpretation of the lensing signal.  It is found that the mean of the $N(z)$ is the most relevant parameter \citep{Amara&Refrigier2007}. Hence we quantify our results in terms of this quantity. As noted above, to attain a precision of order 1\% on the dark energy equation of state parameter, $w$, the mean of the $N(z)$ for each redshift bin must be known to an accuracy of $0.002(1+z)$, and we take this as our nominal requirement.    

As in BLA-10, we first look at the use of a single ``maximum-likelihood'' photo-$z$ for each galaxy.  In Figure-\ref{overall_performance} we plot the r.m.s. dispersion $\sigma_{z}(z)$ and the mean bias $\Delta_{z}(z)$ derived from individual galaxies by comparing the maximum likelihood redshifts with the actual redshifts. The blue line gives the initial performance of the photo-$z$ code, without any removal of outliers. The green line gives the performance after the ``a priori'' identification of outliers (recognized purely from their likelihood curves, see BLA-10 for details) and the red line gives the performance after modifying the likelihood functions to produce the final pdfs, and using the peaks of the pdf. It can be seen that the photo-$z$ performance improves after each of these steps, although it should be noted that almost $20\%$ of galaxies are lost in the cleaning process - most of which would have had a quite usable photo-$z$.  As in BLA-10, the required $\sigma_{z}(z)$ can be achieved relatively easily.

Turning to the more demanding problem of the error on the $N(z)$, the use of maximum likelihood redshifts will not, by definition, include the wings of the $N(z)$ or deal with any catastrophic failures in the photo-$z$.  Figure-\ref{tomo_bin_example} shows a reconstructed tomographic bin for different methods. The true $N(z)$ (shaded histogram) is best represented by the summing of pdfs obtained by applying the L(z) to $\mathscr{L}(z)$ transformation (red line). The mean of each reconstruction is shown by the vertical lines and for the above method the we obtain the minimum bias in the reconstruction. 

In Figure-\ref{deltaz_final} the blue diamonds show the bias in the mean of the $N(z)$ using just the maximum likelihood redshifts after ``a priori'' cleaning. The error bars are simply calculated as the expected uncertainty in the average redshift based on Poisson statistics, knowing $\sigma_{z}(z)$, i.e. we could not detect systematic biases below this level.  The biases using these maximum likelihood redshifts fall well short of the requirement of $0.002(1+z)$ (shown by the shaded region) and are much larger than the Poisson uncertainties in our simulation.  

As in BLA-10, some improvement is obtained by summing the individual $L(z)$ curves to define the $N(z)$, shown as the red squares, but the required performance is only realized when we sum the individual pdfs that are obtained for each galaxy using the procedure outlined above, and described in more detail in BLA-10.  These are shown as black filled circles.  In this case, the error bars come from taking different subsets of the spectroscopic sample to define the L(z) to $\mathscr{L}(z)$ transformation. With this procedure, an uncertainty of $0.002(1+z)$ is met for each of the bins. 

Comparing Figure-\ref{overall_performance} with Figure-2  in BLA-10, it can be seen that there is a degradation of performance in this more realistic simulation compared with that of our earlier paper.  This is clearly the signature of ``template error'' and of the other realistic observational problems that are present in the current study through the use of actual photometric data.  However, the degradation is relatively small: in order to achieve the same performance, about 7$\%$ more galaxies are lost due to ``a priori'' cleaning as compared to BLA-10.  This modest degradation justifies our earlier assumption that template error was unlikely to be a dominant factor, at least with the typical $S/N$ and number of filters expected for foreseeable all-sky photometric surveys.

As described above, the required performance is met with of order 1000 objects to map this transformation.  This number makes sense:  In BLA-10 we pointed out the potential reduction in the number of spectroscopic redshifts needed if these were used to {\it characterize} the photo-$z$ scheme rather than to estimate the $N(z)$ directly.  Our use of spectroscopic redshifts to construct the L(z) to $\mathscr{L}(z)$ mapping transformation is effectively an empirical determination of the errors on the redshifts.   It is not surprising that, to go from a $\sigma_{z} \sim 0.05$ to a bias that is 30 times smaller, requires of order $30^2$ known redshifts, c.f. equation (2) in BLA-10.  This suggests that the number of spectroscopic redshifts would not increase dramatically as the total sample of photo-$z$ objects goes up.

On the other hand, the discussion in Section 2.2 highlights the dangers of using a spectroscopic sample that is not demonstrably complete, because of lingering concerns that the ``catastrophic failure'' rate of the photo-$z$, or even the statistical error $\sigma_{z}$, may be correlated with the success of measuring a spectroscopic redshift, and therefore that the set of galaxies used to set up the scheme may not be representative.  Establishing that the calibrating sample is in fact completely representative, i.e. demonstrating to a possibly skeptical audience that the photo-$z$ performance has been met, will likely require a very much larger sample.

\section{Summary}

This paper is the second to focus on the problem of photo-$z$ performance for weak-lensing cosmology from anticipated all-sky surveys.   In a previous study \citep{Bordoloi09} we used a simulated photometric catalogue that was generated using the same set of templates as were then used to estimate the photo-$z$, thereby removing any issues associated with the choice of templates from the analysis.  

To avoid this potential circularity, we have used in the current study a more realistic photometric catalogue that has been generated by degrading actual photometry of galaxies in the COSMOS field for which secure spectroscopic redshifts are available from zCOSMOS.   In order to generate a simulated photometric catalogue that extends to the required depth in the near-infrared, and to overcome the paucity of spectroscopic redshifts at $z > 1$, we employ the trick of basing the catalogue on bluer photometry of somewhat brighter objects at lower redshifts, in particular simulating an $I_{AB} \leq 24.5$ sample at $0.65 < z < 1.8$ from a $V \leq 22.5$ sample at $0.2 < z < 1.0$.  Such galaxies are well matched in luminosity.  

The photo-$z$ are estimated using synthetic templates with no attempt at optimization, so in this sense no spectroscopic redshifts are formally used for calibration of the photo-$z$. About 10\% of the sample ($\sim 10^{3}$ redshifts) are used in setting up the algorithm that is used to convert the likelihood functions to probability distribution functions, which is essential for the accurate estimate of the mean redshifts.  We suspect that it is the total number and not the percentage which will be relevant if this approach is applied to much larger samples.

We find that, although the performance of the new simulation is slightly worse than the previous, more idealized simulation, the degradation is quite small. In effect, 7\% more galaxies must be eliminated through an ``a priori'' cleaning in order to attain the required performance.  This supports our earlier assertion that ``template error'' would be unlikely to dominate the photo-$z$ performance of surveys with the modest S/N and limited numbers of filters that are likely to be available in all-sky surveys.

The new analysis therefore supports our earlier conclusion that photo-$z$ performance at the level required for precision cosmology are in principle achievable with foreseeable photometric surveys.  While it indicates that a satisfactory photo-$z$ scheme can be set up, it highlights however that large spectroscopic samples of very high completeness and with very reliable spectroscopic redshift determinations may be required in order to actually demonstrate that the required performance has been achieved, because of lingering concerns that redshift incompleteness may be correlated with photo-$z$ accuracy.  This may ultimately be the limiting factor.

\section{Acknowledgements}
This work has been supported by the Swiss National Science Foundation and is based on observations undertaken at the European Southern Observatory (ESO) Very Large Telescope (VLT) under Large Program 175.A-0839. Also based on data collected at Subaru Telescope, which is operated by the National Astronomical Observatory of Japan.

\bibliographystyle{thesis_bibtex}
\bibliography{mybibliography}

\begin{thebibliography}{17}
\expandafter\ifx\csname natexlab\endcsname\relax\def\natexlab#1{#1}\fi

\bibitem[{{Amara} \& {R{\'e}fr{\'e}gier}(2007)}]{Amara&Refrigier2007}
{Amara}, A. \& {R{\'e}fr{\'e}gier}, A. 2007, MNRAS, 381, 1018

\bibitem[{{Bordoloi} {et~al.}(2010){Bordoloi}, {Lilly}, \&
  {Amara}}]{Bordoloi09}
{Bordoloi}, R., {Lilly}, S.~J., \& {Amara}, A. 2010, MNRAS, 406, 881

\bibitem[{{Bruzual} \& {Charlot}(2003)}]{Bruzual&Charlot2003}
{Bruzual}, G. \& {Charlot}, S. 2003, MNRAS, 344, 1000

\bibitem[{{Capak} {et~al.}(2008){Capak}, {Aussel}, {Ajiki}, {McCracken},
  {Mobasher}, {Scoville}, {Shopbell}, {Taniguchi}, {Thompson}, {Tribiano},
  {Sasaki}, {Blain}, {Brusa}, {Carilli}, {Comastri}, {Carollo}, {Cassata},
  {Colbert}, {Ellis}, {Elvis}, {Giavalisco}, {Green}, {Guzzo}, {Hasinger},
  {Ilbert}, {Impey}, {Jahnke}, {Kartaltepe}, {Kneib}, {Koda}, {Koekemoer},
  {Komiyama}, {Leauthaud}, {Lefevre}, {Lilly}, {Liu}, {Massey}, {Miyazaki},
  {Murayama}, {Nagao}, {Peacock}, {Pickles}, {Porciani}, {Renzini}, {Rhodes},
  {Rich}, {Salvato}, {Sanders}, {Scarlata}, {Schiminovich}, {Schinnerer},
  {Scodeggio}, {Sheth}, {Shioya}, {Tasca}, {Taylor}, {Yan}, \&
  {Zamorani}}]{cosmos_photometry_catalogue}
{Capak}, P., {et~al.} 2008, VizieR Online Data Catalog, 2284, 0

\bibitem[{{Daddi} {et~al.}(2007){Daddi}, {Dickinson}, {Morrison}, {Chary},
  {Cimatti}, {Elbaz}, {Frayer}, {Renzini}, {Pope}, {Alexander}, {Bauer},
  {Giavalisco}, {Huynh}, {Kurk}, \& {Mignoli}}]{Daddi2007a}
{Daddi}, E., {et~al.} 2007, ApJ, 670, 156

\bibitem[{{Elbaz} {et~al.}(2007){Elbaz}, {Daddi}, {Le Borgne}, {Dickinson},
  {Alexander}, {Chary}, {Starck}, {Brandt}, {Kitzbichler}, {MacDonald},
  {Nonino}, {Popesso}, {Stern}, \& {Vanzella}}]{Elbaz2007}
{Elbaz}, D., {et~al.} 2007, Astronomy and Astrophysics, 468, 33

\bibitem[{{Feldmann} {et~al.}(2006){Feldmann}, {Carollo}, {Porciani}, {Lilly},
  {Capak}, {Taniguchi}, {Le F{\`e}vre}, {Renzini}, {Scoville}, {Ajiki},
  {Aussel}, {Contini}, {McCracken}, {Mobasher}, {Murayama}, {Sanders},
  {Sasaki}, {Scarlata}, {Scodeggio}, {Shioya}, {Silverman}, {Takahashi},
  {Thompson}, \& {Zamorani}}]{Feldmann2006}
{Feldmann}, R., {et~al.} 2006, MNRAS, 372, 565

\bibitem[{{Hildebrandt} {et~al.}(2008){Hildebrandt}, {Wolf}, \&
  {Ben{\'{\i}}tez}}]{Hildebrandt2008}
{Hildebrandt}, H., {Wolf}, C., \& {Ben{\'{\i}}tez}, N. 2008, Astronomy and
  Astrophysics, 480, 703

\bibitem[{{Hu}(1999)}]{Hu_Tomography}
{Hu}, W. 1999, ApJL, 522, L21

\bibitem[{{Ilbert} {et~al.}(2009){Ilbert}, {Capak}, {Salvato}, {Aussel},
  {McCracken}, {Sanders}, {Scoville}, {Kartaltepe}, {Arnouts}, {Floc'h},
  {Mobasher}, {Taniguchi}, {Lamareille}, {Leauthaud}, {Sasaki}, {Thompson},
  {Zamojski}, {Zamorani}, {Bardelli}, {Bolzonella}, {Bongiorno}, {Brusa},
  {Caputi}, {Carollo}, {Contini}, {Cook}, {Coppa}, {Cucciati}, {de la Torre},
  {de Ravel}, {Franzetti}, {Garilli}, {Hasinger}, {Iovino}, {Kampczyk},
  {Kneib}, {Knobel}, {Kovac}, {LeBorgne}, {LeBrun}, {F{\`e}vre}, {Lilly},
  {Looper}, {Maier}, {Mainieri}, {Mellier}, {Mignoli}, {Murayama}, {Pell{\`o}},
  {Peng}, {P{\'e}rez-Montero}, {Renzini}, {Ricciardelli}, {Schiminovich},
  {Scodeggio}, {Shioya}, {Silverman}, {Surace}, {Tanaka}, {Tasca}, {Tresse},
  {Vergani}, \& {Zucca}}]{2009ApJ...690.1236I}
{Ilbert}, O., {et~al.} 2009, ApJ, 690, 1236

\bibitem[{{Kitzbichler} \& {White}(2007)}]{kitzbichler&White2007}
{Kitzbichler}, M.~G. \& {White}, S.~D.~M. 2007, MNRAS, 376, 2

\bibitem[{{Lilly} {et~al.}(1996){Lilly}, {Le Fevre}, {Hammer}, \&
  {Crampton}}]{Lilly1996}
{Lilly}, S.~J., {Le Fevre}, O., {Hammer}, F., \& {Crampton}, D. 1996, ApJL,
  460, L1+

\bibitem[{{Lilly} {et~al.}(2007){Lilly}, {Le F{\`e}vre}, {Renzini}, {Zamorani},
  {Scodeggio}, {Contini}, {Carollo}, {Hasinger}, {Kneib}, {Iovino}, {Le Brun},
  {Maier}, {Mainieri}, {Mignoli}, {Silverman}, {Tasca}, {Bolzonella},
  {Bongiorno}, {Bottini}, {Capak}, {Caputi}, {Cimatti}, {Cucciati}, {Daddi},
  {Feldmann}, {Franzetti}, {Garilli}, {Guzzo}, {Ilbert}, {Kampczyk}, {Kovac},
  {Lamareille}, {Leauthaud}, {Borgne}, {McCracken}, {Marinoni}, {Pello},
  {Ricciardelli}, {Scarlata}, {Vergani}, {Sanders}, {Schinnerer}, {Scoville},
  {Taniguchi}, {Arnouts}, {Aussel}, {Bardelli}, {Brusa}, {Cappi}, {Ciliegi},
  {Finoguenov}, {Foucaud}, {Franceschini}, {Halliday}, {Impey}, {Knobel},
  {Koekemoer}, {Kurk}, {Maccagni}, {Maddox}, {Marano}, {Marconi}, {Meneux},
  {Mobasher}, {Moreau}, {Peacock}, {Porciani}, {Pozzetti}, {Scaramella},
  {Schiminovich}, {Shopbell}, {Smail}, {Thompson}, {Tresse}, {Vettolani},
  {Zanichelli}, \& {Zucca}}]{Lilly2007}
{Lilly}, S.~J., {et~al.} 2007, ApJS, 172, 70

\bibitem[{{Lilly} {et~al.}(2009){Lilly}, {Le Brun}, {Maier}, {Mainieri},
  {Mignoli}, {Scodeggio}, {Zamorani}, {Carollo}, {Contini}, {Kneib}, {Le
  F{\`e}vre}, {Renzini}, {Bardelli}, {Bolzonella}, {Bongiorno}, {Caputi},
  {Coppa}, {Cucciati}, {de la Torre}, {de Ravel}, {Franzetti}, {Garilli},
  {Iovino}, {Kampczyk}, {Kovac}, {Knobel}, {Lamareille}, {Le Borgne}, {Pello},
  {Peng}, {P{\'e}rez-Montero}, {Ricciardelli}, {Silverman}, {Tanaka}, {Tasca},
  {Tresse}, {Vergani}, {Zucca}, {Ilbert}, {Salvato}, {Oesch}, {Abbas},
  {Bottini}, {Capak}, {Cappi}, {Cassata}, {Cimatti}, {Elvis}, {Fumana},
  {Guzzo}, {Hasinger}, {Koekemoer}, {Leauthaud}, {Maccagni}, {Marinoni},
  {McCracken}, {Memeo}, {Meneux}, {Porciani}, {Pozzetti}, {Sanders},
  {Scaramella}, {Scarlata}, {Scoville}, {Shopbell}, \&
  {Taniguchi}}]{Lilly2009_article}
{Lilly}, S.~J., {et~al.} 2009, ApJS, 184, 218

\bibitem[{{Madau}(1995)}]{Madau1995}
{Madau}, P. 1995, ApJ, 441, 18

\bibitem[{{Mannucci} {et~al.}(2010){Mannucci}, {Cresci}, {Maiolino}, {Marconi},
  \& {Gnerucci}}]{Mannucci2010}
{Mannucci}, F., {Cresci}, G., {Maiolino}, R., {Marconi}, A., \& {Gnerucci}, A.
  2010, MNRAS, 408, 2115

\bibitem[{{Refregier} {et~al.}(2010){Refregier}, {Amara}, {Kitching}, {Rassat},
  {Scaramella}, {Weller}, \& {Euclid Imaging Consortium}}]{EIC2010}
{Refregier}, A., {Amara}, A., {Kitching}, T.~D., {Rassat}, A., {Scaramella},
  R., {Weller}, J., \& {Euclid Imaging Consortium}, f.~t. 2010, ArXiv
  e-prints:-1001.0061

\end{thebibliography}
\end{document}